\documentstyle[10pt]{article}
\makeatletter
\newdimen\normalarrayskip              % skip between lines
\newdimen\minarrayskip                 % minimal skip between lines
\normalarrayskip\baselineskip
\minarrayskip\jot
\newif\ifold             \oldtrue            
\def\arraymode{\ifold\relax\else\displaystyle\fi} % mode of array enrties
\def\eqnumphantom{\phantom{(\theequation)}}     % right phantom in eqnarray
\def\@arrayskip{\ifold\baselineskip\z@\lineskip\z@
     \else
     \baselineskip\minarrayskip\lineskip2\minarrayskip\fi}
\def\@arrayclassz{\ifcase \@lastchclass \@acolampacol \or
\@ampacol \or \or \or \@addamp \or
   \@acolampacol \or \@firstampfalse \@acol \fi
\edef\@preamble{\@preamble
  \ifcase \@chnum
     \hfil$\relax\arraymode\@sharp$\hfil
     \or $\relax\arraymode\@sharp$\hfil
     \or \hfil$\relax\arraymode\@sharp$\fi}}
\def\@array[#1]#2{\setbox\@arstrutbox=\hbox{\vrule
     height\arraystretch \ht\strutbox
     depth\arraystretch \dp\strutbox
     width\z@}\@mkpream{#2}\edef\@preamble{\halign \noexpand\@halignto
\bgroup \tabskip\z@ \@arstrut \@preamble \tabskip\z@ \cr}%
\let\@startpbox\@@startpbox \let\@endpbox\@@endpbox
  \if #1t\vtop \else \if#1b\vbox \else \vcenter \fi\fi
  \bgroup \let\par\relax
  \let\@sharp##\let\protect\relax
  \@arrayskip\@preamble}
\def\eqnarray{\stepcounter{equation}%
              \let\@currentlabel=\theequation
              \global\@eqnswtrue
              \global\@eqcnt\z@
              \tabskip\@centering
              \let\\=\@eqncr
              $$%
 \halign to \displaywidth\bgroup
    \eqnumphantom\@eqnsel\hskip\@centering
    $\displaystyle \tabskip\z@ {##}$%
    &\global\@eqcnt\@ne \hskip 2\arraycolsep
         $\displaystyle\arraymode{##}$\hfil
    &\global\@eqcnt\tw@ \hskip 2\arraycolsep
         $\displaystyle\tabskip\z@{##}$\hfil
         \tabskip\@centering
    &{##}\tabskip\z@\cr}
\makeatother

\def\beq{\begin{equation}}
\def\eeq{\end{equation}}
\def\bea{\begin{eqnarray}}
\def\eea{\end{eqnarray}}
\def\stackreb#1#2{\mathrel{\mathop{#2}\limits_{#1}}}

\begin{document}

\begin{titlepage}
\begin{center}
{\it P.N.Lebedev Institute preprint} \hfill {FIAN/TD-14/93}\\
{\it I.E.Tamm Theory Department} \hfill hepth@xxx/???????
\begin{flushright}{September 1993}\end{flushright}
\vspace{0.1in}{\Large\bf Quantization of string theory for $c \leq 1$ }\\[.4in]
{\large  S. Kharchev, A.Marshakov
\footnote{talk given by A.M. at the
$XXVII$th International Symposium on Elementary particle physics,
Wendisch-Rietz, Germany (September 1993)
}}\\
\bigskip {\it  P.N.Lebedev Physics
Institute \\ Leninsky prospect, 53, Moscow, 117 924, Russia,
\footnote{E-mail address:
tdparticle@glas.apc.org \ \  mars@td.fian.free.net}
}

\end{center}
\bigskip \bigskip

\begin{abstract}
We consider the canonical quantization scheme for $c \leq 1$ ($(p,q)$ -) string
theories
and compare it with what is known from matrix model approach. We derive
explicitly a trivial ($\equiv $ topological) solution. We discuss a ``dressing"
operator which in principle allows one to obtain a non-trivial solution, but
an explicit computation runs into a
problem of analytic continuation of the formal expressions for
$\tau $-functions.
We discuss also the application of proposed scheme to the case of discrete
matrix model and consider some parallels with mirror symmetry and background
independence in string theory.
\end{abstract}

\end{titlepage}

\newpage
\setcounter{footnote}0

\section{Introduction}

Recent development of string theory brought us to a to a more or less sensible
progress in
{\it explicit} description of the exact solutions to the simplest string
models.
Fortunately enough, it turns out
that
the correlation functions in such theories
\footnote{which are usually called topological $2d $ gravity plus topological
matter, though this is not too informative and complete definition}
can be determined through their
generation functions which not only solve the well-known integrable equations
of the KP (Toda lattice) hierarchy but also appear to be a particular solution
having simple integral representation form. This allows to find a kind of
(rather degenerate) field
theory representations for these solutions and try to interpret these
representations as a second-quantized or string field theory.

{}From a general outlook we have here an example of a new phenomenon in the
study of quantum systems -- their relation to the
solution of {\it classical} integrable equations.
Among other examples one can mention the cases when correlation functions in
the {\it quantum} integrable systems are
solutions to the {\it classical} integrable equations \cite{japanese,IIKS}.
In string theory we have
\footnote{and looking for in more complicated cases}
slightly different statement: generating function for
correlators in some (exotic) physical system is a $\tau $-function
\footnote{or Baker-Akhiezer function if normalized (see below)} of KP or
Toda-lattice hierarchy.

{}From mathematical point of view this ''integrable structure" is related to
the underlying
module space structures whose topological characteristics can be computed as
correlation functions in a corresponding topological theory. Moreover,
almost all what is known about higher-dimensional (topological) theories
relies upon corresponding structures of module spaces of the target-space
theory. However, it is not still clear how to derive integrable structure
directly from the properties of corresponding module spaces. Below we will
review ``integrable" approach to the case of simplest $2d$ gravity theories:
$(p,q)$ models interacting with $2d$ gravity and try to clarify their features
which should allow one to find parallels with the less trivial
higher-dimensional cases.

It is necessary to point out that we are still dealing with two different
problems in string theory. Starting from the end, the second one is related
with the properties of integrable systems describing (or hypothetically
describing) string theory. The success in $c \leq 1$ case here is mostly
related to the fact that one needs to work with the simplest hierarchy of
integral equations (one-component KP or Toda-lattice) where lots of useful
facts were known in advance. The less trivial cases (multicomponent
hierarchies, hierarchies with non-commuting flows {\it etc}) are either much
less known or even not considered at all in the literature. However, the
first problem is even more fundamental - what is the sense of the integrable
equations from the first principles of string theory. At the moment, there
exists a lot of different approaches (or languages) more or less useful when
understanding this or that group of facts but neither of them gives complete
understanding of the sense of the equations in the space of coupling
constants.

\section{String equation and Heisenberg algebra}

First, we will review the problem in the form related to the search of
solutions to non-perturbative gravity interacting with the $(p,q)$-matter,
which
still exists in a not satisfactory form except for trivial $(p,1)$
``topological" case (see
\cite{LGGKM,KM1,KM2}) where it can be represented in terms of matrix model in
external matrix field.

It is well-known \cite{Douglas}, that it can be reduced to a problem of
description of a particular representation of the Heisenberg algebra.
Indeed,
consider the representation of the Heisenberg operators, satisfying the string
equation
$[\hat P,\hat Q] = 1$  in the ``momentum" space

\begin{equation}\label{repr}
\hat P =  \lambda
$$
$$
\hat Q = {\partial \over \partial \lambda } + Q(\lambda )
\end{equation}

{}From the point of view of the KP hierarchy, we will also add some additional
requirements on the ``spectral parameter" implying that

\begin{equation}
\lambda  = \mu ^p
\end{equation}
then $(p,q)$ models correspond to the case where  $Q(\lambda )$  should be a
{\it polynomial} of $\mu $ of degree $q$ \cite{FKN1}, (while
the corresponding wave functions should have specific asymptotics when  $\mu
\rightarrow  \infty $).

Wave functions  of this problem appear to be the Baker-Akhiezer functions
of the corresponding integrable system and
when acting on wave functions conditions (\ref{repr}) get the form of the
Kac-Schwarz equations \cite{KS,S}:

\begin{equation}\label{KS}
\lambda \varphi _i(\mu ) = \sum  _j W_{ij}\varphi _j(\mu )
$$
$$
\hat A\varphi _i(\mu ) = \sum  _j A_{ij}\varphi _j(\mu )
\end{equation}
where

\begin{equation}
\lambda  = W(\mu ) \sim  \mu ^p
$$
$$
A^{(W,Q)} \equiv  N^{(W,Q)}(\mu ){1\over W'(\mu )}
{\partial \over \partial \mu } [N^{(W,Q)}(\mu )]^{-1} =
$$
$$
= {1\over W'(\mu )} {\partial \over \partial \mu } - {1\over 2}
{W''(\mu )\over W'(\mu )^2} + Q(\mu )
\end{equation}

The standard way to construct wave functions of the theory is to
define the Fock vacuum by

\begin{equation}\label{Fock}
\hat A\Psi _0= 0
\end{equation}
with an obvious solution

\begin{equation}
\Psi _0 = \sqrt{W'(\mu)}\exp{\int{QdW}}
\end{equation}
and the corresponding $\tau $-function is a determinant
projection of  higher states

\begin{equation}
\Psi _n \sim W^n \Psi _0
\end{equation}
to the states with a {\it canonical} asymptotics

\begin{equation}
\varphi _i(\mu )
\stackreb{\mu \rightarrow \infty }{\rightarrow} \mu ^{i-1}
\end{equation}
forming the conventional basis in the space of wave functions - the point of
infinite-dimensional Grassmannian.

The only simple case arises when the Kac-Schwarz equations
(\ref{KS}) have trivial solution, {\it i.e.} when  $p=1$.
Starting from normalization
$\varphi _1(\mu ) = 1$ (corresponding to $\Psi_0 = \exp{\int{Qd\mu}}$), and
using first of eqs.(\ref{KS}) one can always get
$\Psi_n = \mu^n\exp{\int{Qd\mu}} \rightarrow \varphi _i(\mu ) = \mu ^{i-1}$
{\it exactly}. Then the second condition of
(\ref{KS}) is fulfilled {\it automatically} for {\it any} $Q(\mu )$.

However, one can see that the corresponding solutions are not ``physically"
trivial
in the sense that they are related to more meaningful solutions
by a kind of Fourier
transformation. Indeed, it has been observed
\cite{KM1,KM2} that the system of equations (\ref{KS}) posseses a {\it duality}
symmetry which relates $(p,q)$ to $(q,p)$ solution.
The duality transformation for the Baker-Akhiezer functions looks like

\begin{equation}
\psi ^{(P,Q)}(z) = [P'(z)]^{1/2}\int   dQ\ e^{P(z)Q(x)}\psi ^{(Q,P)}(x)
[Q'(x)]^{-1/2}
\end{equation}
and it can be also written for the basis vectors in the Grassmannian

\beq\label{dual}
\phi _i(\mu ) = [W'(\mu )]^{1/2} \exp ( - \left.S_{W,Q}\right |
_{x=\mu })  \int   d{\cal M}_Q(x)f_i(x) \exp \ S_{W,Q}(x,\mu )
\eeq
with

\begin{equation}\label{act}
d{\cal M}_Q(x) = dx \sqrt{Q'(x)}
$$
$$
S_{W,Q}(x, \mu) = - \int_{}^{x}{WdQ} + W(\mu)Q(x)
\end{equation}
and for the partition functions

\beq\label{dualpf}
\tau^{(W,Q)}\,\,[M] =
$$
$$
= C[V,M]\int DX\tau^{(Q,W)}\,\,[X]\exp \left \{Tr[1/2 \log Q'(X) +
\int_{M}^{X}W(z)dQ(z) + W(M)Q(X)] \right \}
\eeq
(here, better to consider {\it normalized} partition function
$\tau ^{(W,Q)} \rightarrow Z^{(W,Q)} \rightarrow
\Psi _{BA}^{(W,Q)}(t_k - {1 \over k}Tr M^{-k}$).
It makes possible to obtain solutions for
nontrivial models -- topological $(p,1)$
models \cite{KMMMZ} and their Landau-Ginzburg deformations \cite{LGGKM}.

\begin{equation}\label{LanGin}
\varphi _i(\mu ) = (p\mu ^{p-1})^{1/2}\exp (- \sum t_k\mu ^k)\int   dx\ x^{i-1}
\exp (-V(x) + x\mu ^p)
\end{equation}
which are {\it dual}  to $(p,1)$ model in the above sense.

Here, we immediately run into a puzzle: how to interpret this from the point
of view of quantization theory. Indeed, the duality transformation (\ref{dual})
is nothing but a transformation from $\hat p$ to $\hat q$ quantization
procedure or from one to another representation of quantum algebra and as it
is well-known the quantization should be independent of this. From this
point of view $(p,1)$ and $(1,p)$ or trivial theory should be equivalent.
On the other hand we know that the partition functions for $(p,1)$ theories
are nontrivial and correspond to some well-known topological theories (twisted
$N=2$ Landau-Ginzburg theories) interacting with topological gravity ($(2,1)$
model corresponds to pure topological gravity and generates intersection
indices on module spaces of Riemann surfaces with punctures). Thus, there
should be a way to extract all this ``topologcal" information from a ``dual"
partition function $\tau ^{(1,p)} \equiv 1$. We will return to this problem
below.

\section{Flows in the space of solutions}

Now let us discuss briefly how one can try to construct explicitly less
trivial solutions.
First, let us mention that the corresponding projection

\beq
Pr: \Psi _{n-1} \rightarrow \varphi _n
\eeq
will be highly nontrivial (singular) in this case, though the corresponding
basis vectors $\Psi _n$ span the whole space.

Another option is to start from above integral formulas (see also \cite{KM2}).
One can try to construct more or less explicit representation for complicated
solutions starting from known ones and using a sort of ``dressing" operator.
This way is certainly not very useful from technical point of view and does not
lead us to a final result, but it demonstrates some valuable properties of
these
solutions and in particular gives some observations close to that of
\cite{Losev}.

The main formula can be derived from a simple fact that some sort of
``dressing transformations" just maps one solution of (\ref{KS}) to another
\footnote{One can easily recognize in this formula an element of $W_{\infty}$
in the sense of \cite{FKN2}}

\begin{equation}\label{dress}
\varphi _i(z|t) = \exp{\sum{(t_k - {p \over p+1}\delta _{k,p+1})z^k}}
\exp{\sum{C_{ij} z^{pi}\hat{A}^j}}=
$$
$$
= (pz^{p-1})^{1/2}\exp (- \sum t_kz^k)
\exp \ \sum C_{ij}\lambda ^i({\partial \over \partial \lambda })^j\int
dx\ x^{i-1} \exp (- {x^{p+1}\over p+1} + x\lambda )
\end{equation}

This formula has a simple quantum mechanical interpretation. From the point of
view of the Heisenberg algebra representation it can be
considered as a quantum mechanical matrix element of
the following form

\begin{equation}
\langle q_1 | \exp{H(p,q)} |q_0 \rangle
\end{equation}
with

\begin{equation}
H(p,q) = \sum{C_{ij} p^iq^j}
\end{equation}
Thus, we have obtained a sort of path integral representation for a nontrivial
solution (instead of a trivial solution in the topological case). An
interesting question is if it can be computed via localization or other
technique and does it have a solution for finite-dimensional matrices
$C_{ij}(t)$.

\subsection{Perturbation expansion}

Now we are going to present few explicit examples of the formula (\ref{dress}),
computed in perturbation theory. First, let us start with
``expansion" of the matrix $ C_{ij}$ into different pieces which would have
different physical meaning

\begin{equation}\label{expansion}
\sum \ C_{ij}x^j \equiv  \Omega_i (x)
\end{equation}
We are going to consider the contribution of the two first terns in the
sum (\ref{expansion}), taking them in the following form

\begin{equation}
\Omega _0(x) = -V(x) + {x^{p+1}\over p+1} = - \sum_{j = 1}^{p}{{v_j \over
j}x^j}
$$
$$
\Omega _1(x) = {1 \over F'(x)} = \epsilon x^q
\end{equation}
One can demonstrate that the first one corresponds to the Landau-Ginzburg
deformation of the potential (\ref{LanGin}) while the second one brings to
a reparameterization in the space of fields and gives a nontrivial co-ordinate
$ Q(X)$ in the sense of \cite{Losev}.

Indeed

\begin{equation}
\exp{C_{0j}\left( {\partial \over \partial\lambda} \right)^j} \int{dx x^{i-1}
\exp{\left( - {x^{p+1} \over p+1} + \lambda x \right)}}=
$$
$$
= \int{dx x^{i-1} \exp{\left( - {x^{p+1} \over p+1} + \Omega _0 (x) +
\lambda x \right)}} = \int{dx x^{i-1} \exp{\left(- V(x) + \lambda x \right)}}
\end{equation}
what corresponds to (\ref{LanGin}), and

\begin{equation}
\exp{\left( \Omega _0( {\partial \over \partial\lambda} )+ \lambda\Omega_1(
{\partial \over \partial\lambda} ) \right)}
\int{dx x^{i-1} \exp{\left( - {x^{p+1} \over p+1} + \lambda x \right)}} =
$$
$$
\int{dx x^{i-1}
\exp{\left( - {x^{p+1} \over p+1} \right)}}
\exp{\left( \Omega_0(x) + \Omega_1(x){\partial \over
\partial x }\right)}
\exp{\lambda x}
\end{equation}
Then, using that

\beq
\exp {\left( \Omega_0(x) + \Omega_1(x) {\partial \over \partial x} \right)} =
 \exp{B} \exp {\Omega_1 \partial}
\eeq
with

\beq
B = \int_x^{F^{-1}\left(1+F(x)\right)}{d\xi \Omega_0(\xi) \over \Omega_1(\xi)}
$$
$$
F'(x) \equiv {1 \over \Omega_1(x)}
\eeq
and choosing $F(x) = {x^{1-q} \over \epsilon (1-q)}$,
$\Omega_0 \rightarrow \Omega _0 + {\epsilon \over 2} x^{q-1}$, one gets

\beq
\exp{\left( \Omega_0(x) + \Omega_1(x){\partial \over
\partial x }\right)} \exp{\lambda x} =
$$
$$
= \exp {B} \exp \lambda Q(x)
\sim \sqrt{1+\epsilon x^{q-1}}
\exp \lambda Q(x)
\eeq
where

\beq
Q(x) = \left( x^{1-q} + \epsilon (1-q) \right)^{1 \over {1-q}}
= x + \epsilon x^q + ...
\eeq
We see that this is indeed an infinitesimal reparameterization of ``spectral
co-ordinate" $X \rightarrow
Q(X) = X + \epsilon X^q + ...$.

\subsection{Analytic continuation. Example of discrete matrix model}

So, we see that the formula (\ref{dress}) is consistent with what we know about
the exact solutions {\it perturbatively}. It means that one can consider
non-trivial solutions as perturbations over trivial topological ones. However,
the problems runs into the difficulties of analytic continuation and the most
trivial example to demonstrate this is a discrete matrix model \cite{KMMM}.

Indeed, when computing the integral

\beq\label{dmm}
\int DH \exp \left( - Tr \sum t_k H^k \right)
\eeq
there exists an essential difference between
$ t_k = {1 \over 2} \delta _{k,2} $ and
``double-scaling" cases, considered in \cite{MMMM} (see also references
therein).

In the first case, due to special properties of the Hermite polynomials
(with measure ${1\over 2}H^2$) which have rather simple integral representation
there exists a GKM-like {\it determinant} formula which means that this case
is actually quite similiar to the {\it topological} $(p,1)$
models. The matrix integral computed in such a way gives a generation function
for the correlators only in the {\it Gaussian} discrete matrix model, and that
means that the Gaussian matrix model is an example of the simplest topological
theory.

In contrast, considered ``double-scaling" limits can be
obtained from (\ref{dmm}) only as a highly-nontrivial analytic continuation.
For
example, formulae from \cite{KMMM} look like

\beq
\sum_{n>2, i_k>0}
C_{i_3 ... i_n}(t_2)
t_3^{i_3}...t_n^{i_n}
\eeq
where $C_{i_3 ... i_n}(t_2)$ are in general {\it non-analytic} functions of
$t_2$, like

\beq
\int DH e^{- t_2 Tr H^2} =
\int \prod _{i,j} dH_{ij} e^{-t_2 \sum _{i,j}H_{ij}H_{ji}}
\sim t_2^{-{N^2\over 2}}
\eeq
The partitions functions for generic $2d$ gravity models will be
in contrast {\it non-analytic} functions of higher times, for example
$t_4$, $t_6$ {\it etc}.

\subsection{BA-function computation}

Let us compute the quasiclassical expansion for the Baker-Akhiezer function.
We start with the $(p,1)$ case.

\beq\label{ba}
\Psi (z,t) = \sqrt{pz^{p-1}}\int dx \exp \left( - V(x) + xz^p \right)=
$$
$$
= \sqrt{pz^{p-1}}\exp \left( - V(\mu) + \mu z^p \right)
\int dx \exp \left( - {1\over 2}V''(\mu)(x-\mu )^2
- {1\over 3!}V'''(\mu )(x-\mu )^3 - ... \right)=
$$
$$
= \sqrt{pz^{p-1}\over W'(\mu )}\exp \left( - V(\mu) + \mu z^p \right)
\int d\xi \exp \left( - {1\over 2}\xi ^2
- {1\over 3!}{W''(\mu )\over W'(\mu )^{3/2}}\xi ^3 - ...
- {1\over n!}{W^{(n-1)}(\mu )\over W'(\mu )^{n/2}}\xi ^n - ...\right)
\eeq
The integral results in

\beq
\langle \exp \sum y_k \xi^k \rangle = \langle \sum P_m(y) \xi ^m \rangle
= \sum P_{2m}(y) \langle \xi ^{2m} \rangle
\eeq
with

\beq
y_1 = y_2 = 0
$$
$$
y_n = - {1\over n!}{W^{(n-1)}(\mu )\over W'(\mu )^{n/2}}, \ \ n>2
$$
$$
\langle ... \rangle \equiv \int d\xi \exp \left(- {1\over 2} \xi ^2 \right) ...
\eeq
Computing gaussian integrals one gets

\beq
\langle \xi ^{2m} \rangle = (2m-1)!!
\eeq
and using that

\beq
V(\mu ) - \mu W(\mu ) = - \sum ^{p+1}_{-\infty} t_kz^k
$$
$$
\mu = {1\over p}\sum ^{p+1}_{-\infty} kt_kz^{k-p}
$$
$$
{1\over W'(\mu )} = {1\over p^2} \sum ^{p+1}_{-\infty} k(k-p)t_kz^{k-2p}
\eeq
we get

\beq
\Psi (z,t) = \sqrt{pz^{p-1}\over W'(\mu )}
\exp \left( -\sum t_k z^k \right)
\left[ {5\over 12} {W''^2\over W'^3} - {1\over 8}{W'''\over W'^2} + ...\right]
\eeq
The terms in the square brackets behaive like ${1\over z^{p+1}}$ and
they are {\it
dependent} on the first $p$ terms given by the expansion of the pre-factor,
giving the only interesting contribution

\beq\label{BAtop1}
\Psi (z,t) = \sqrt{\sum _{k = -\infty}^{p+1} {k(k-p)\over p^2}t_k z^{k-p-1}}
\exp \left( -\sum t_k z^k \right) + ...
\eeq

The Baker-Akhiezer functions relation:

\beq\label{BAtop2}
\Psi (\tilde t, \mu ) = \exp \sum \tilde t_k \mu ^k =
\exp \sum t_k z^k
\eeq
where first sum is {\it finite} while the second one - {\it infinite}. The
set of times $\{ \tilde t \}$ is simply related with the coefficients of the
``superpotential" $W(\mu ) = \sum _{j = 0}^p v_j \mu ^j$

\beq
\tilde t_k = - {1\over k(1-k)} Res \mu ^{1-k} dW(\mu ) = {1\over k} v_j
\delta _{j,k+1}
\eeq

\section{Mirror symmetry and pq-duality}

Now let us briefly make some comments on relation between the quantization
scheme we advocated above for $c \leq 1$ theories and a popular question of
mirror symmetry
(see for example \cite{Vafa,Witten} and references therein). One may hope
that the paralells we will discuss below can shed light on possible appearance
of the integrable hierarchies in the ``higher-dimensional" theories.

First, the simple observation is that the quasiclassics distinguishes $p$ and
$q$ from a symmetric formulation - indeed the classical limit {\it depends} on
what we call an area operator in the theory($\beta_+ = \sqrt{{2p \over q}}$
or $\beta_- = \sqrt{{2q \over p}}$). In such case we have a sort of
mirror map between $(p,q)$ and $(q,p)$ theory similiar to
$R \rightarrow {1\over R}$ in the case of $c = 1$ theories.
In general, if we have two mirror manifolds only quantum
theories are equivalent (not classical ones which depend on configuration
space) and two different classical limits correspond to different spaces.
It is well known already when studying classical limit of corresponding $2d$
conformal field theories, one has to fix either $p > q$ or $p < q$ in the
$(p,q)$ model and only one ``screening" survives in the classical limit.
\footnote{The most simple and illustrative example is the same effect in the
WZNW model where only one screening has nice classical interpretation
\cite{GMMOS}}.

In the classical limit instead of $[\hat P,\hat Q] = 1$ we have the
Poisson bracket

\beq\label{PB}
\{ W,Q\} = 1
\eeq
which is actually generated by

\beq
\{ z,t_1\} = 1
$$
$$
\{\tilde z, \tilde t_1\} = 1
\eeq
(where $z^p = W(\mu )$ and $\tilde z^q = Q(\mu )$), see \cite{Kri,TakTak} etc.
For trivial $(1,p)$ topological theories $\tilde z = \mu $.

{}From this point of view what we consider is a quantization of a symplectic
manifold

\beq
\omega = \delta W \wedge \delta Q
\eeq
(in the simplest case)

\beq
\omega = \delta z \wedge \delta t_1
\eeq
and we can consider it along the lines \cite{Witten}.

The corresponding connection is ``action" (\ref{act})

\beq
S = \int WdQ + S_0
$$
$$
dS = \delta W \wedge \delta Q
\eeq
and $S_0$ parameterizes an ``initial point". Now, it is obvious that in
the proposed quantization scheme the set of coupling constants depends
on the way of quantization, so does the solutions (potentials) of the
hierarchy, $\tau $- or the BA function {\it etc}. This can be easily seen
already on the example of (\ref{BAtop1}) and (\ref{BAtop2}).

\section{Conclusion}

In these notes we have tried to present some ideas how the results of
formulation of non-perturbative string theory in terms of hierarchies
of integrable equations can appear through ``canonical" way of quantization.
One can hope, that this way will bring us to more understanding of the problem
what is second-quantized string theory and why does this quantum theory possess
a structure of hierarchies of classical integrable equations.

However, this ``canonical" way of quantization is still very far from being
completed. First, even in the simplest case of $c \leq 1$ theories we do not
have detailed explanation of the geometry underlying the Heisenberg string
equation. The fact that a generation function for correlators in string theory
appears as a wave function in the second-quantized theory gives an analogy
with a similiar effect in the Chern-Simons theory.

The main problem is still how to generalize the language of integrable
hierarchies to the higher-dimensional theories(like topological strings
in the Calabi-Yau backgrounds, critical strings etc). The basic moment
for the integrable hierarchies is the appearance of spectral curve via
Miwa transformation. The dependence of the partition function of ``Miwa times"
should be identified with the dependence of the partition function of
higher-dimensional theory on the ``homology co-ordinate", or put differently
on the co-ordinate in the tangent bundle to the module space. Then, the
``classical" component of the time variables - an element of the
finite-dimensional small phase space should be identified with the
co-ordinate on the module space itself.

Such way of thinking immediately leads us to an idea that the second-quantized
string theory should be based on the {\it quantum module space}. This object
naturally appears in the frames of the Chern-Simons theory (see
\cite{FockRosly} and references therein). This problem certainly deserves
further investigation.

Let us finally add few comments about holomorphic anomaly. The
``quasiclassical" $\tau $-function obeys a homogeneous relation

\beq
\sum t_j {\partial \over \partial t_j } \log \tau _0 = 2 \log \tau _0
\eeq
spoilt by the contribution of the one-loop correction, having the form,
for example, for the $(2,1)$ theory

\beq\label{holan}
\sum t_j {\partial \over \partial t_j } \log \tau  - 2 \log \tau =
- {1\over 24}
\eeq
The similiar expressions appear when one considers the logariphm of the
partition function for the higher-dimensional theories \cite{holan} and this
should mean that the expression (\ref{holan}) should have a similiar nature.

There is certainly a lot of other open questions. We are going to return to
them in a separate publication.

\section{Acknowledgements}
We would like to thank V.Fock, A.Gerasimov, A.Losev, N.Nekrasov and B.Voronov
for very useful discussions.
The work was in part supported by Russian Fund for Fundamental
Research and by a grant of American Physical Society.

\end{document}